\documentclass[twocolumn]{aastex62}

\graphicspath{{./}{figures/}}

\received{}
\revised{}
\accepted{}
\submitjournal{ApJ}

\shorttitle{The weakening outburst of V582 Aur}
\shortauthors{Zsidi et al.}

\begin{document}

\title{The weakening outburst of the young eruptive star V582 Aur}

\correspondingauthor{Gabriella Zsidi}
\email{zsidi.gabriella@csfk.mta.hu}

\author{G. Zsidi}
\affiliation{Konkoly Observatory, Research Centre for Astronomy and Earth Sciences, Hungarian Academy of Sciences, \\
Konkoly-Thege Mikl\'os \'ut 15-17, Budapest, Hungary}
\affiliation{E\"otv\"os Lor\'and University, Department of Astronomy, Budapest, Hungary}

\author{P. \'Abrah\'am}
\affiliation{Konkoly Observatory, Research Centre for Astronomy and Earth Sciences, Hungarian Academy of Sciences, \\
Konkoly-Thege Mikl\'os \'ut 15-17, Budapest, Hungary}

\author{J. A. Acosta-Pulido}
\affiliation{Instituto de Astrof\'\i{}sica de Canarias, Avenida V\'\i{}a L\'actea, Tenerife, Spain}
\affiliation{Departamento de Astrof\'\i{}sica, Universidad de La Laguna, Tenerife, Spain}

\author{\'A. K\'osp\'al}
\affiliation{Konkoly Observatory, Research Centre for Astronomy and Earth Sciences, Hungarian Academy of Sciences, \\
Konkoly-Thege Mikl\'os \'ut 15-17, Budapest, Hungary}
\affiliation{Max Planck Institute for Astronomy, Heidelberg, Germany}

\author{M. Kun}
\affiliation{Konkoly Observatory, Research Centre for Astronomy and Earth Sciences, Hungarian Academy of Sciences, \\
Konkoly-Thege Mikl\'os \'ut 15-17, Budapest, Hungary}

\author{Zs. M. Szab\'o}
\affiliation{Konkoly Observatory, Research Centre for Astronomy and Earth Sciences, Hungarian Academy of Sciences, \\
Konkoly-Thege Mikl\'os \'ut 15-17, Budapest, Hungary}
\affiliation{E\"otv\"os Lor\'and University, Department of Astronomy, Budapest, Hungary}

\author{A. B\'odi}
\affiliation{Konkoly Observatory, Research Centre for Astronomy and Earth Sciences, Hungarian Academy of Sciences, \\
Konkoly-Thege Mikl\'os \'ut 15-17, Budapest, Hungary}
\affiliation{MTA CSFK Lend\"ulet Near-Field Cosmology Research Group, Budapest, Hungary}

\author{B. Cseh}
\affiliation{Konkoly Observatory, Research Centre for Astronomy and Earth Sciences, Hungarian Academy of Sciences, \\
Konkoly-Thege Mikl\'os \'ut 15-17, Budapest, Hungary}

\author{N. Castro Segura}
\affiliation{Departamento de Astrof\'\i{}sica, Universidad de La Laguna, Tenerife, Spain}
\affiliation{Physics and Astronomy Department, University of Southampton, Southampton, UK}

\author{O. Hanyecz}
\affiliation{Konkoly Observatory, Research Centre for Astronomy and Earth Sciences, Hungarian Academy of Sciences, \\
Konkoly-Thege Mikl\'os \'ut 15-17, Budapest, Hungary}

\author{B. Ign\'acz}
\affiliation{Konkoly Observatory, Research Centre for Astronomy and Earth Sciences, Hungarian Academy of Sciences, \\
Konkoly-Thege Mikl\'os \'ut 15-17, Budapest, Hungary}
\affiliation{E\"otv\"os Lor\'and University, Department of Astronomy, Budapest, Hungary}

\author{Cs. Kalup}
\affiliation{Konkoly Observatory, Research Centre for Astronomy and Earth Sciences, Hungarian Academy of Sciences, \\
Konkoly-Thege Mikl\'os \'ut 15-17, Budapest, Hungary}
\affiliation{E\"otv\"os Lor\'and University, Department of Astronomy, Budapest, Hungary}

\author{L. Kriskovics}
\affiliation{Konkoly Observatory, Research Centre for Astronomy and Earth Sciences, Hungarian Academy of Sciences, \\
Konkoly-Thege Mikl\'os \'ut 15-17, Budapest, Hungary}

\author{L. M\'esz\'aros}
\affiliation{Konkoly Observatory, Research Centre for Astronomy and Earth Sciences, Hungarian Academy of Sciences, \\
Konkoly-Thege Mikl\'os \'ut 15-17, Budapest, Hungary}

\author{A. Ordasi}
\affiliation{Konkoly Observatory, Research Centre for Astronomy and Earth Sciences, Hungarian Academy of Sciences, \\
Konkoly-Thege Mikl\'os \'ut 15-17, Budapest, Hungary}

\author{A. P\'al}
\affiliation{Konkoly Observatory, Research Centre for Astronomy and Earth Sciences, Hungarian Academy of Sciences, \\
Konkoly-Thege Mikl\'os \'ut 15-17, Budapest, Hungary}

\author{K. S\'arneczky}
\affiliation{Konkoly Observatory, Research Centre for Astronomy and Earth Sciences, Hungarian Academy of Sciences, \\
Konkoly-Thege Mikl\'os \'ut 15-17, Budapest, Hungary}

\author{B. Seli}
\affiliation{Konkoly Observatory, Research Centre for Astronomy and Earth Sciences, Hungarian Academy of Sciences, \\
Konkoly-Thege Mikl\'os \'ut 15-17, Budapest, Hungary}
\affiliation{E\"otv\"os Lor\'and University, Department of Astronomy, Budapest, Hungary}

\author{\'A. S\'odor}
\affiliation{Konkoly Observatory, Research Centre for Astronomy and Earth Sciences, Hungarian Academy of Sciences, \\
Konkoly-Thege Mikl\'os \'ut 15-17, Budapest, Hungary}

\author{R. Szak\'ats}
\affiliation{Konkoly Observatory, Research Centre for Astronomy and Earth Sciences, Hungarian Academy of Sciences, \\
Konkoly-Thege Mikl\'os \'ut 15-17, Budapest, Hungary}



\begin{abstract}

V582 Aur is a pre-main sequence FU Orionis type eruptive star, which entered a brightness minimum in 2016 March due to changes in the line-of-sight extinction. Here, we present and analyze new optical $B$, $V$, $R_C$ and $I_C$ band multiepoch observations and new near-infrared $J$, $H$ and $K_S$ band photometric measurements from 2018 January--2019 February, as well as publicly available mid-infrared WISE data. We found that the source shows a significant optical--near-infrared variability, and the current brightness minimum has not completely finished yet. If the present dimming originates from the same orbiting dust clump that caused a similar brightness variation in 2012, than our results suggest a viscous spreading of the dust particles along the orbit. Another scenario is that the current minimum is caused by a dust structure, that is entering and leaving the inner part of the system. The WISE measurements could be consistent with this scenario. Our long-term data, as well as an accretion disk modeling hint at a general fading of V582 Aur, suggesting that the source will reach the quiescent level in $\sim$80 years.

\end{abstract}



\section{Introduction}\label{sec:intro}

V582 Aur is an FU Orionis (FUor) type young eruptive star. These systems show decade long optical--infrared outbursts caused by temporary increase in the mass accretion rate from the inner circumstellar disk onto the star \citep{hk96}. While in some cases the initial brightening has been observationally documented, other objects were classified as FUors based on their spectral characteristics \citep{connelley2018}. Typically the initial brightening is followed by a longer fading phase. For example FU Ori has been in outburst for 80 years, and V1057 Cyg for 50 years \citep{audard2014b}. So far only one FUor, V346 Normae seemed to finish its outburst, after 30 years \citep{kraus2016}.

\begin{deluxetable*}{lcllllcccl}
\tablecaption{Optical and near-infrared photometry in magnitudes for
	  V582~Aur. Numbers in parentheses give the formal uncertainty of the last digit. \label{tab:phot}}
\tablehead{
\colhead{Date} & \colhead{JD$\,{-}\,$2,450,000} & \colhead{$B$} & \colhead{$V$} & \colhead{$R$} & \colhead{$I_C$} & \colhead{$J$} & \colhead{$H$} & \colhead{$K_S$} & \colhead{Telescope}
}
\startdata
1998 Oct 08 & 1094.92 &  &  &  &  &  9.62(2) &  8.49(2) &  7.74(2) & 2MASS \\
\hline
2018 Jan 14 & 8133.31 & 17.83(3) & 15.95(2) & 14.70(2) & 13.22(1) &  &  &  & Schmidt \\
2018 Jan 25 & 8144.30 & 17.61(2) & 15.68(2) & 14.46(2) & 13.02(1) &  &  &  & Schmidt \\
2018 Feb 06 & 8156.41 & 17.56(3) & 15.74(2) & 14.53(1) & 13.05(1) &  &  &  & Schmidt \\
2018 Feb 18 & 8168.28 & 17.73(7) & 15.96(1) & 14.69(1) & 13.24(1) &  &  &  & Schmidt \\
2018 Feb 25 & 8175.29 & 18.18(4) & 15.98(7) & 14.77(1) & 13.26(2) &  &  &  & Schmidt \\
2018 Mar 08 & 8186.31 & 18.04(4) & 16.11(1) & 14.86(1) & 13.34(1) &  &  &  & Schmidt \\
2018 Mar 10 & 8188.36 & 18.00(6) & 16.08(3) & 14.85(1) & 13.32(1) &  &  &  & Schmidt \\
2018 Mar 14 & 8192.26 & 17.92(3) & 16.15(3) & 14.86(1) & 13.35(1) &  &  &  & Schmidt \\
2018 Apr 02 & 8211.28 & 17.83(1) & 16.10(3) & 14.85(1) & 13.35(1) &  &  &  & Schmidt \\
2018 Apr 10 & 8219.30 & 18.00(24) & 16.36(9) & 14.92(9) & 13.47(4) &  &  &  & Schmidt \\
2018 Apr 11 & 8220.29 & 18.07(7) & 16.27(2) & 14.99(1) & 13.48(1) &  &  &  & Schmidt \\
2018 Apr 13 & 8222.31 & 18.13(4) & 16.26(4) & 15.08(2) & 13.53(1) &  &  &  & Schmidt \\
2018 Apr 14 & 8223.30 & 18.27(2) & 16.37(1) & 15.10(1) & 13.57(1) &  &  &  & Schmidt \\
2018 Apr 15 & 8223.86 &  &  &  & &  & 9.66(7) & & Liverpool \\
2018 Apr 18 & 8227.30 & 18.17(5) & 16.42(2) & 15.16(1) & 13.61(1) &  &  &  & Schmidt \\
2018 Apr 21 & 8230.31 & 18.20(5) & 16.42(2) & 15.12(1) & 13.58(1) &  &  &  & Schmidt \\
2018 Aug 29 & 8360.60 & 17.20(9) & 15.54(2) & 14.35(2) & 12.93(2) &  &  &  & Schmidt \\
2018 Aug 31 & 8362.55 & 17.29(6) & 15.50(1) & 14.31(1) & 12.93(1) &  &  &  & Schmidt \\
2018 Sep 15 & 8377.59 & 17.25(3) & 15.49(1) & 14.33(2) & 12.93(1) &  &  &  & Schmidt \\
2018 Sep 21 & 8383.46 & 17.35(5) & 15.53(3) & 14.35(2) & 12.92(1) &  &  &  & Schmidt \\
2018 Sep 24 & 8386.49 & 17.35(9) & 15.45(2) & 14.37(2) & 12.92(1) &  &  &  & Schmidt \\
2018 Sep 26 & 8387.56 & 17.39(7) & 15.47(2) & 14.33(2) & 12.92(1) &  &  &  & Schmidt \\
2018 Sep 28 & 8389.56 & 17.22(5) & 15.51(1) & 14.32(1) & 12.94(1) &  &  &  & Schmidt \\
2018 Oct 24 & 8416.47 & 17.91(9) & 16.11(1) & 14.94(2) & 13.47(1) &  &  &  & Schmidt \\
2018 Oct 31 & 8422.62 & 17.69(3) & 15.91(2) & 14.72(1) & 13.29(1) &  &  &  & Schmidt \\
2018 Nov 01 & 8423.64 & 17.80(7) & 15.94(5) & 14.69(2) & 13.24(1) &  &  &  & Schmidt \\
2018 Nov 05 & 8427.64 & 17.53(3) & 15.78(1) & 14.58(2) & 13.14(1) &  &  &  & Schmidt \\
2018 Nov 07 & 8429.50 & 17.51(4) & 15.69(1) & 14.52(1) & 13.08(1) &  &  &  & Schmidt \\
2018 Nov 10 & 8432.67 & & & & & 10.62(1) & 9.40(1) & 8.60(1) & TCS \\
2018 Nov 12 & 8434.64 & 17.42(3) & 15.66(2) & 14.44(2) & 13.02(1) &  &  &  & Schmidt \\
2018 Dec 05 & 8458.39 & 17.34(1) & 15.59(1) & 14.41(1) & 13.00(1) &  &  &  & Schmidt \\
2018 Dec 14 & 8466.56 & 17.45(2) & 15.60(2) & 14.41(1) & 13.01(1) &  &  &  & Schmidt \\
2018 Dec 24 & 8477.38 & 17.61(4) & 15.75(3) & 14.55(2) & 13.11(1) &  &  &  & Schmidt \\
2019 Jan 07 & 8491.31 & 17.62(6) & 15.75(1) & 14.57(1) & 13.11(1) &  &  &  & Schmidt \\
2019 Feb 05 & 8520.42 & 17.33(4) & 15.54(1) & 14.37(1) & 12.96(1) &  &  &  & Schmidt \\
\enddata
\end{deluxetable*}


The physical origin of the outburst is still debated, for a review of the proposed mechanisms we refer to \cite{audard2014b}. 
During an outburst, a considerable amount of matter falls from the inner disk onto the star, and this matter is then replenished by gas and dust transported inward from the outer regions. The outburst heat may evaporate dust particles and rearrange the density structure close to the star \citep[e. g.][]{kun2011}. These structural changes may occur on the dynamical timescale of the inner disk, months to years, and may cause variations in the line-of-sight extinction towards the central part of the system.

   \begin{figure*}[t!]
   \centering
   \includegraphics[height=\textwidth, angle=90]{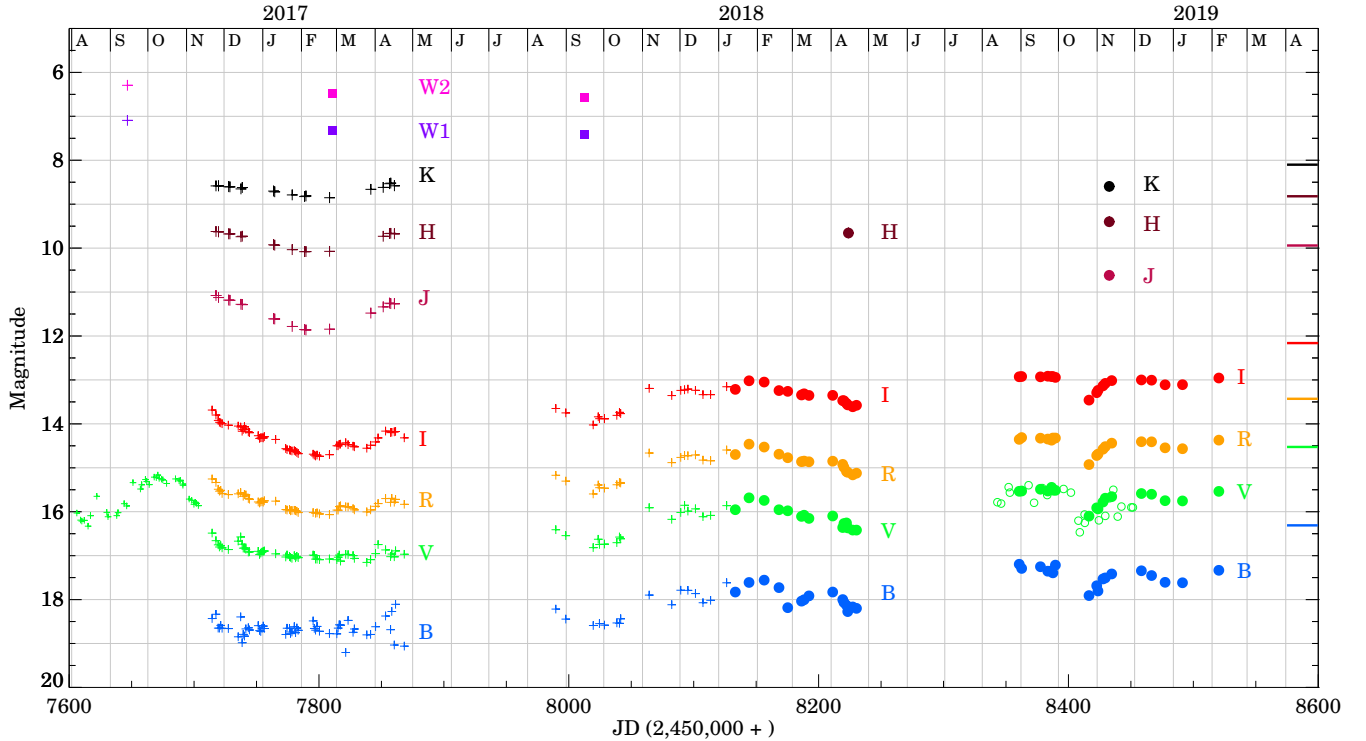}
      \caption{Light curves of V582 Aur. Filled circles represent data from this work, plus signs are from A18, and empty circles are the latest data from the ASAS-SN sky survey \citep{shappee2014,kochanek2017}. The new mid-infrared WISE data points are marked with squares. Tick marks on the top indicate the first day of each month. The measured errors are smaller than  or equal to the symbol size. On the right side we show the maximal brightness levels from 2014 with horizontal lines.}
         \label{fig:light_curve}
   \end{figure*}

V582 Aur has been in outburst since 1985 \citep{semkov2013}. In our previous paper \citep[hereafter A18]{abraham2018} we analyzed the long term multiwavelength photometric variability of V582 Aur, which we interpreted as a combined effect of changing accretion and obscuration. In A18, we investigated the evolution of the outburst, focusing on two dimming events. We found that the brightness minima are related to changing line-of-sight extinction, and speculated that a dust clump is orbiting in the inner disk with the period of 5 years. The second dimming event started in 2016, and has not finished yet. In this paper we present new observations in order to follow the further evolution of V582 Aur and confront the new data with the extinction scenario of A18.

\section{Observations}\label{sec2}

We performed $B$, $V$, $R_C$, and $I_C$ band photometric measurements at Konkoly Observatory (Hungary), using the 60/90/180 cm Schmidt telescope. The observations were carried out between 2018 January 14 and 2019 February 5. We also obtained near-infrared $J$, $H$, and $K_S$ band images using the Telescopio Carlos Sanchez (TCS) at the Teide Observatory (Spain) on 2018 Nov 10. For the data reduction of both the Schmidt and TCS measurements, we followed the same procedure as described in A18. In addition, we performed a single $H$ band measurement on 2018 April 15 using the 2 m Liverpool telescope at the Observatorio del Roque de Los Muchachos, Canary Islands, Spain (Prop. No. CQ18A01, PI: J. A. Acosta-Pulido). For this observation we used the IO:I camera built on a 2048x2048 HAWAII 2RG chip \citep{barnsley2016}. We obtained images at 9 dither positions, with 5.8 s exposure time each, and applied aperture photometry. The calibration and error calculus were based on the 2MASS magnitudes of 3 nearby bright stars. All photometric results are presented in Table \ref{tab:phot}. 

Mid-infrared photometry at 3.4 $\mu$m (W1) and 4.6 $\mu$m (W2), obtained by the WISE space telescope  \citep{wright2010}, was presented in A18. Here, we complement these sequences by two new data points from 2017. For these epochs we collected all time-resolved observations from the NEOWISE-R
Single Exposure Source Table, and computed their average and standard deviation. As in A18, we added in quadrature 2.4\% and 2.8\% to the formal uncertainty of the average value, in order to account for the uncertainty of the absolute calibration in the W1 and W2 bands, respectively. All available WISE measurements are included in Table~\ref{tab:wise}.


\section{Results}\label{sec3}

\subsection{Optical--near-infrared light curves}

   \begin{figure*}[ht!]
   \includegraphics[width=\textwidth]{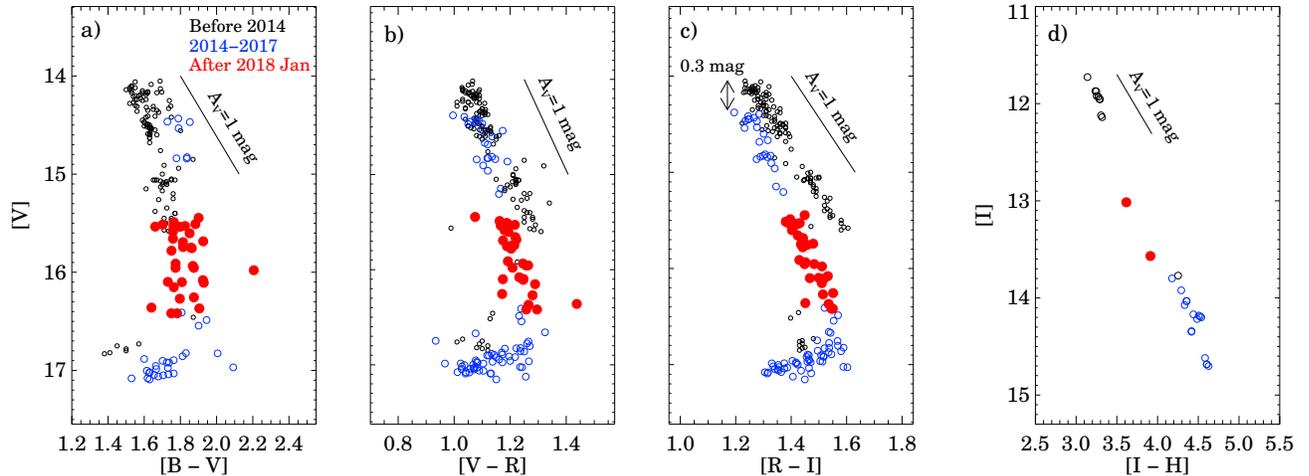}
   \caption{Color--magnitude diagrams of V582 Aur. It is similar to Fig. 10 in A18, but we corrected a mistake in the numbering of the y-axis. The filled red circles indicate our new measurements (Table \ref{tab:phot}), which fill the brightness gap between the previous data. The data points from 2014--2017 (published in A18) are overplotted with blue open circles, and the earlier data before 2014 are drawn as black circles. The lines correspond to an extinction change of $A_V=1$ mag.}
   \label{fig:color}
   \end{figure*}


Our new photometric results are plotted in Figure \ref{fig:light_curve}. For reference we also plotted the earlier light curves from A18 until 2018 January 7. Following the minimum of 2017 February, the optical light curves show a brightening trend until 2018 February. In A18, we expected that this trend would continue and the source would reach the maximum brightness level (measured in 2013--2014, see horizontal lines in Fig.~\ref{fig:light_curve}) during 2018. However, our new optical data indicate a break in this trend. After reaching a peak in 2018 February, the source started dimming again. The latest observations from 2018 August--2019 February show similar brightness level to the beginning of the year, interrupted by a local minimum in 2018 October. Currently the source is as bright as it was in 2018 August--September, but is still about 1 magnitude fainter than it was in 2013--2014.

After 2018 February, the shapes of the $B$, $V$, $R_C$, and $I_C$ light curves are more similar to each other than in 2016/2017 (Fig.~\ref{fig:light_curve}). Our near-infrared $H$~band photometric point on 2018 Apr 15 shows the same brightness level as the last near-infrared data point in 2017 April. On 2018 Nov 10 the source was significantly brighter in the $J$ and $H$~bands than in 2017, while no change in its $K_S$ brightness was detected. These magnitudes are however still fainter by 0.5--0.7 mag than the maximum brightness levels before 2016 (see A18). Our photometric results imply that the dimming of V582 Aur which started in 2016 has not finished yet, and its length is at least 3 years, significantly exceeding the length of the 2012 dimming.

\subsection{Color changes}

In Figure \ref{fig:color}, we show the color-magnitude diagrams based on our optical and near-infrared data. We plotted data points measured before 2014 (black circles), photometry from 2014--2017 (blue circles), and our new observations from Table \ref{tab:phot} (filled red circles). The earlier data points did not populate the brightness range between $V$=15.5 and 16.5 mag. The new data points after 2018 January fill this brightness gap, representing an intermediate brightness level. Figure~\ref{fig:color}$a$-$c$ suggest that the data points obtained before 2014 and the data points measured in 2014--2017 occupy different locations in the color-magnitude space. Comparing the colors of the two samples at any given $V$ band magnitude, the earlier data (black circles) are bluer in Fig.~\ref{fig:color}$a$, have similar [$V-R$] colors in Fig.~\ref{fig:color}$b$, and are systematically redder in Fig.~\ref{fig:color}$c$ than the 2014--2017 measurements. Moreover, there is an additional discrepancy in the distribution of points before and after 2014. They exhibit similar but parallel patterns in such a way that the data after 2014 seem to be $\sim$0.3 mag dimmer in $V$ band for the same color. In Fig.~\ref{fig:color}$a$-$c$, our new photometry after 2018 January remarkably match the color-magnitude trends outlined by the data points from 2014--2017. 

Fig.~\ref{fig:color}$d$, which shows an $I_C$ vs. [$I_C-H$] diagram, indicates that the new data points seem to match the trend outlined by the 2014--2017 measurements. Extrapolating from the 2014--2017 data to brighter $I_C$ magnitudes, there is a hint that the data points before 2014 were systematically redder at a given $I$~band magnitude than the later photometry, similarly to Fig.\ref{fig:color}$c$.

\subsection{WISE measurements}
\label{sec:wise}

\begin{figure*}[ht!]
\includegraphics[height=\textwidth, angle=90]{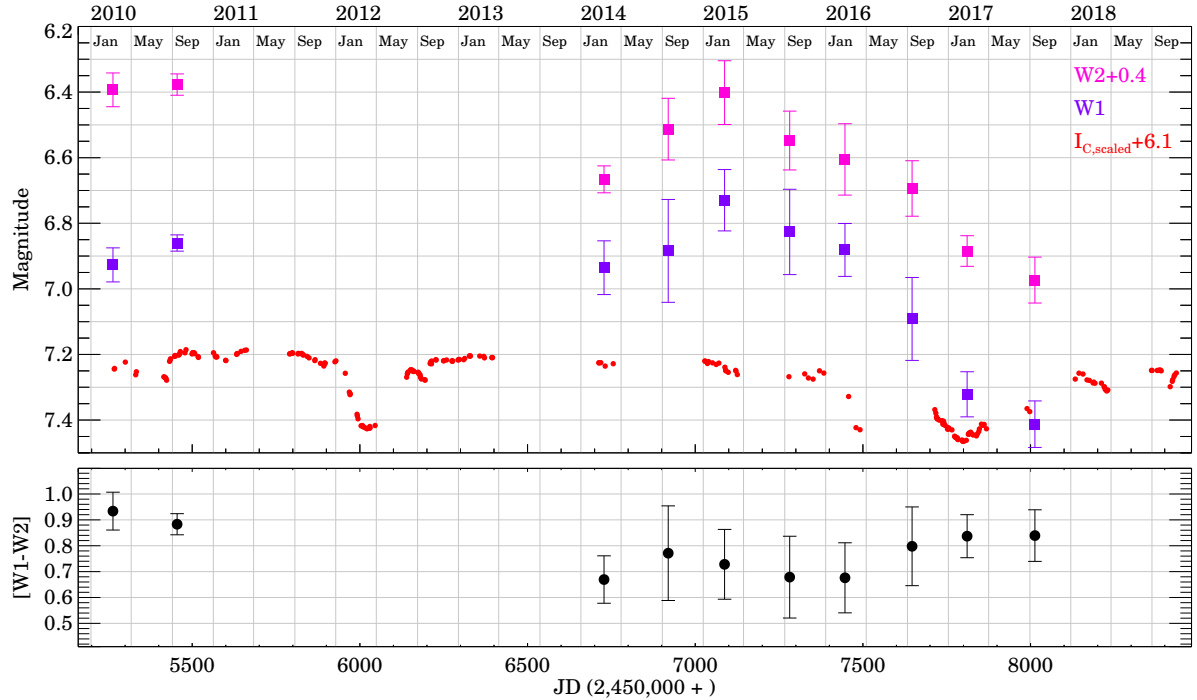}
\caption{The upper panel shows the near-infrared light curves of V582 Aur. The squares show the WISE measurements and the red circles indicate the  $I_C$-band data scaled down with the Cardelli reddening law and shifted along the y-axis (see Sec.~\ref{sec:wise}). The WISE color change during the observing period is indicated on the lower panel.}
\label{fig:wise}
\end{figure*}


All available WISE measurements from Table~\ref{tab:wise} are plotted in Figure~\ref{fig:wise}. The first two measurements during 2010 revealed a slight brightening of the source. By the beginning of 2014, V582~Aur was fainter again, and the drop of brightness with respect to 2010 was significantly larger in the W2 than in the W1~band. During 2014, the WISE light curves show a brightening trend, and after reaching a maximum in 2015 March both light curves started a linear fading phase, which is still followed by the source.

\begin{deluxetable}{lcll}
\tablecaption{Mid-infrared WISE data in magnitudes. Numbers in parentheses give the uncertainty of the last digit. \label{tab:wise}}
\tablehead{
\colhead{Date} & \colhead{JD$\,{-}\,$2,450,000} & \colhead{W1} & \colhead{W1}
}
\startdata
2010 Mar 08 & 5263.624 & 6.93(5) & 5.99(5) \\
2010 Sep 15 & 5454.918 & 6.86(2) & 5.98(3) \\
2014 Mar 11 & 6728.090 & 6.94(8) & 6.27(4) \\
2014 Sep 19 & 6919.520 & 6.88(16) & 6.11(9) \\
2015 Mar 05 & 7087.476 & 6.73(9) & 6.00(9) \\
2015 Sep 16 & 7281.657 & 6.83(13) & 6.15(9) \\
2016 Feb 28 & 7446.604 & 6.88(8) & 6.21(11) \\
2016 Sep 15 & 7646.648 & 7.09(13) & 6.29(8) \\
2017 Feb 26 & 7810.799 & 7.32(7) & 6.48(5) \\
2017 Sep 16 & 8012.973 & 7.41(7) & 6.57(7) \\
\enddata
\end{deluxetable}

In A18, we demonstrated that the optical and near-infrared variability of V582 Aur could be, to a large part explained by changing extinction in the line-of-sight. In order to check if the mid-infrared variability, measured by WISE, was also due to extinction variations, we compared the mid-infrared light curves with the optical ones. We overplotted in Fig. \ref{fig:wise} a modified version of the $I_C$ light curve scaled down by the relative amplitudes of the interstellar extinction at the respective wavelengths (0.79 $\mu$m and 3.4 $\mu$m). This scaled down light curve would be the expected mid-infrared behavior if the light variations were caused by the same extinction at both the optical and near-infrared wavelengths. Interestingly, the measured WISE light curves do not follow the expected trend, suggesting a different physical mechanism as the origin of the variation. The [W1$-$W2] color variations during the observed period, plotted in the bottom panel of Figure \ref{fig:wise}, show that the source was bluer in early 2014 and early 2016, while it was redder in 2017, although not as red as in 2010.

\subsection{New Gaia distance}

In A18, following \citet{kun2017},  we assumed a distance of 1.32 kpc for V582~Aur. The Gaia DR2 parallax, however, implies a larger distance of 2.4$_{-0.4}^{+0.7}$~kpc \citep{bailer-jones2018}. Note that V582 Aur seems to be associated with a complex of bright rim clouds \citep[see Fig. 3 in][]{kun2017}. The closest OB star, thus the most likely ionization source is HD 281147, whose Gaia DR2 distance is $2.4_{-0.2}^{+0.3}$~kpc. This result also supports the new distance value for V582 Aur. We also checked the Gaia DR2 distances for a list of stars used to define the Auriga OB1 and Auriga OB2 associations by \citet{humphreys1978}. We found that the distances for the two associations overlap, although Aur~OB1 ($d = 1.9 \pm 0.7$ kpc) seems to be closer to the Sun than Aur OB2 ($d = 2.8 \pm 0.5$ kpc). Based on these results, it is unclear which association V582 Aur belongs to. Adopting the distance of 2.4 kpc for the FUor, several fundamental parameters would change. The luminosity and accretion rate would increase to 500--1050 L$_{\odot}$ and 8$\times 10^{-5}$ M$_{\odot}$/yr respectively, making the object perhaps the most luminous FUor  in the list of \cite{audard2014b}. The mass of the circumstellar material derived in A18 would also change to 0.1 M$_{\odot}$.

\section{Discussion}

\subsection{Nature of the obscuring material}

In A18, we demonstrated that the multiwavelength brightness variations of V582 Aur could be explained by variable extinction towards the source. We explored two possibilities depending on whether the eclipses in 2012 and in 2016--2018 were caused by the same orbiting dust clump, or by two independent dust structures. Based on the first scenario, we speculated about an elongated dust clump orbiting the star at a radius of 2.8 au, having a length of $\sim$3.5 au, and a mass of 0.004 $M_{\oplus}$. Our new results show that the variability detected after 2018 February also matches the previous color-magnitude trends (Fig. \ref{fig:color}), implying that changing extinction is still the dominant physical mechanism behind the light variations. However, just this fact does not help us decide between the two possibilities. In the following, we discuss how our observations are consistent with one or the other scenario.

In the light of our new data, the 2016--2018 minimum significantly differs in length and shape from the one in 2012. Therefore, if both fading events were caused by the same dust clump, as we supposed in A18, then the clump covers a more extended part of the orbit in 2016--2018 than in 2012. Since the length of the current minimum is at least 2 years, the clump could be stretched over $\sim$7 au if adopting an orbital radius of 2.8 au from A18. It suggests a viscous spreading of the dust particles along the orbit. Since the WISE measurements do not follow the scaled down $I_C$-band light curve (Fig. \ref{fig:wise}), either the orbiting dust clump does not cover that part of the inner disk where the mid-infrared radiation is emitted or the clump itself contributes to the mid-infrared radiation.

The other explanation for the observed dimming events includes individual dust clouds entering and leaving the inner part of the system as giant exocomets. In this picture the inward moving dust structure might pass the line-of-sight and may be responsible for the optical dimming events. Approaching the central object, the dust warms up, increasing the mid-infrared thermal emission of the system and making the observed [W1--W2] WISE color bluer. We speculate that this may be the reason of the mid-infrared brightening of V582 Aur in 2014 and early 2015 (Fig. \ref{fig:wise}). When the dust structure is leaving the system, it cools down and emits less at the mid-infrared WISE wavelengths. We may see the beginning of this effect in the latest WISE measurements.

The light curves and the color evolution of V582 Aur show striking similarity to those of an other well known young stellar object, RW Aur A \citep[and the references therein]{dodin2018}. The physical reason for the behavior of RW Aur A is still debated. The possible mechanisms, like dusty disk wind \citep{petrov2015,bozhinova2016}, or a warped inner disk \citep{facchini2016} might be invoked to explain the variability of V582 Aur as well.

\subsection{Prediction for the end of the outburst}
\label{sec:prediction}

The color-magnitude diagram (Figure \ref{fig:color}) indicates that the distribution of points before and after 2014 outlined similar patterns, however the later dataset have $\sim$0.3 mag fainter values in the $V$-band. In A18, we identified the origin of the pattern as a combination of the interstellar extinction and the UXor blueing effect. 
UXors, a subclass of Herbig Ae/Be stars named after the prototype object UX Orionis, are defined by their irregular deep brightness minima in the light curve. As they start fading, their colors become redder in the color-magnitude diagram, however after reaching a certain faintness, they exhibit a color reversal. This is the so-called blueing effect \citep[][and references therein]{the1994, natta2000}. The most widely accepted physical explanation for this phenomenon assumes a system seen almost edge-on. As a dust clump begins obscuring the star, the observed light becomes redder according to the dust extinction law. When the star is completely obscured by the dust clump, only the bluer scattered light from the circumstellar environment remains, causing a color reversal in the color-magnitude diagram. In Figure \ref{fig:color}, the upper linear part of the pattern is parallel to the extinction path shown by lines in the panels. The lower horizontal part is probably the signature that the central object is significantly obscured and the remaining emission is dominated by scattered light. Here, we speculate that the $\sim$0.3 mag difference arises from the absolute fading of the source, reflecting the long-term evolution of the outburst. In this scenario, the 0.3 mag fading during $\sim$5 years would mean that the source will reach the quiescent level in approximately 80 years. This decades long trend is characteristic for FUor type stars. Future observations will show whether the experienced color change continues. The light curves of the WISE measurements differs significantly from those at the optical wavelengths, suggesting different underlying phenomena.

\subsection{Modeling with an accretion disk}
\label{sec:accdisk}

 \begin{figure}[ht!]
   \includegraphics[width=\columnwidth]{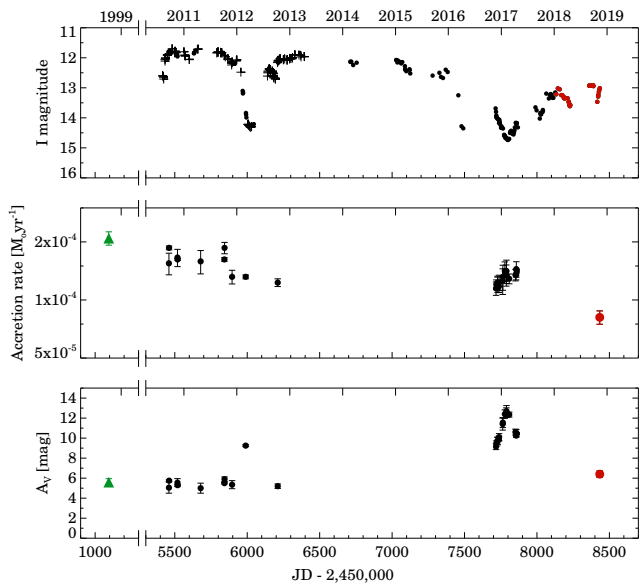}
   \caption{Time evolution of the accretion rate (middle panel) and the
extinction (bottom panel) derived from our accretion disk model (Sect. \ref{sec:accdisk}). The top panel shows the I-band light curve. Black points are from A18, red symbols are from this work, green symbols indicate results based on data from the 2MASS catalog.}
   \label{fig:accdisk}
   \end{figure}

In A18, we fitted the near-infrared spectral energy distributions at different epochs with a simple accretion disk model. We adopted a steady, optically thick, and geometrically thin viscous accretion disk, with a radially constant mass accretion rate. We found that until 2017 April, the mass accretion rate was approximately constant, while the line-of-sight extinction episodically increased during the dimming periods in 2012 and in 2016--2018. Here, we repeat the modeling using the 2.4 kpc distance measured by Gaia, and also including our new $J$, $H$, and $K_S$ measurements from 2018 Nov 10 and the 2MASS measurements from 1998 October 8. Our results are plotted on Fig. \ref{fig:accdisk}. The comparison of this figure and Fig. 11 in A18 shows that the model provided higher accretion rates due to the larger distance of 2.4 kpc. The line-of-sight extinction also increased in the new calculations, since the higher accretion rate corresponds to hotter disk. This requires higher line-of-sight extinction in order to reproduce our measurements. For the new epoch in 2018 we derived an extinction of $A_V\simeq 6.5$~mag, which is significantly lower than what we obtained for the deepest minimum in 2017 February and is only slightly higher than the values measured outside the dimmings, in the maximum brightness state. The corresponding accretion rate value in 2018 November, however, is lower by a factor of 2 than was in the maximum brightness state before 2017. Comparing these results with earlier ones, we found that the extinction in 1998 was on the same level as in 2010-2013, apart from the dimming in 2012. The accretion rate, however, was then higher than at any time later. The long term decreasing trend in the accretion rate is consistent with our previous conclusion on the long-term fading of the source (Sect. \ref{sec:prediction}).

\subsection{Future prospects with Gaia and TESS}

Our results indicate that the light curves of V582 Aur show the combined effect of extinction related dimming events and a general long term decay of the outburst. The latter trend predicts that the source will never return to the pre-2014 state and the current brightness levels may be already close to an unobscured state with a lower central luminosity. Future observations can verify this conclusion and constrain the full length of the eruption. For instance, the ongoing Gaia sky survey will provide improved photometry in two optical passbands starting from 2014. Using the Gaia $G_{RP}$ band observations, which is close to the $I_C$ band (see Fig. 3 in \citealt{jordi2010}), we will be able to not only compare our ground-based measurements with the high precision Gaia photometric results, but we can also examine the long-term variability of V582 Aur. During the nominal 5-year mission, Gaia will provide approximately 50 photometric points for V582 Aur, which will cover a large part or the whole ongoing dimming event. Furthermore, the Transiting Exoplanet Survey Satellite (TESS) will also carry out observations in the direction of Auriga, providing 30 minutes cadence photometry in a broad band, which is centered on the effective wavelength of the $I_C$ band (see Fig. 1 in \citealt{ricker2015}). Considering V582 Aur's ecliptic latitude, TESS will probably provide a 28-day long uninterrupted light curve for this target. These high cadence measurements can reveal the short-term or small amplitude variations, which are invisible for ground-based telescopes and may reveal the fine structure of the obscuring dust cloud.


\acknowledgments

This project has received funding from the European Research Council (ERC) under the European Union's Horizon 2020 research and innovation programme under grant agreement No 716155 (SACCRED). This work was supported by the Lend\"ulet grant LP2012-31 of the Hungarian Academy of Sciences. This project has been supported by the GINOP-2.3.2-15-2016-00003 grant of the Hungarian National Research, Development and Innovation Office (NKFIH). The Liverpool Telescope is operated on the island of La Palma by Liverpool John Moores University in the Spanish Observatorio del Roque de los Muchachos of the Instituto de Astrofisica de Canarias with financial support from the UK Science and Technology Facilities Council. 

%

\vspace{5mm}

\end{document}